Reconsidering the relationship of the El Niño–Southern Oscillation and the Indian monsoon using ensembles in Earth system models

Mátyás Herein[1,2,*], Gábor Drótos[2,3,4], Tamás Bódai[5], Frank Lunkeit[1], Valerio Lucarini[1,5,6]

[1]CEN, Meteorological Institute, University of Hamburg, Hamburg, Germany

[2]MTA–ELTE Theoretical Physics Research Group, and Institute for Theoretical Physics, Eötvös University, Budapest, Hungary

[3]Instituto de Física Interdisciplinar y Sistemas Complejos, CSIC-UIB, Palma de Mallorca, Spain

[4]Max-Planck-Institut für Meteorologie, Hamburg, Germany

[5]Centre for the Mathematics of the Planet Earth, Department of Mathematics and Statistics, University of Reading, Reading, UK

[6]Walker Institute for Climate System Research, University of Reading, Reading, UK

*hereinm@gmail.com

**Abstract**

We study the relationship between the El Niño–Southern Oscillation (ENSO) and the Indian summer monsoon in ensemble simulations from state-of-the-art climate models, the Max Planck Institute Earth System Model (MPI-ESM) and the Community Earth System Model (CESM). We consider two simple variables: the Tahiti–Darwin sea-level pressure difference and the Northern Indian precipitation. We utilize ensembles converged to the system's snapshot attractor for analyzing possible changes (i) in the teleconnection between the fluctuations of the two variables, and (ii) in their climatic means. (i) With very high confidence, we detect an increase in the strength of the teleconnection, as a response to the forcing, in the MPI-ESM under historical forcing between 1890 and 2005, concentrated to the end of this period. We explain that our finding does not contradict instrumental observations, since their existing analyses regarding the nonstationarity of the teleconnection are either methodologically unreliable, or consider an ill-defined teleconnection concept. In the MPI-ESM we cannot reject stationarity between 2006 and 2099 under the Representative Concentration Pathway 8.5 (RCP8.5), and in a 110-year-long 1-percent pure $CO_2$ scenario; neither can we in the CESM between 1960 and 2100 with historical forcing and RCP8.5. (ii) In the latter ensembles, the climatic mean is strongly displaced in the phase space projection spanned by the two variables. This displacement is nevertheless linear. However, the slope exhibits a strong seasonality, falsifying a hypothesis of a universal, emergent relationship between these two climatic means, excluding applicability in an emergent constraint.

**1. Introduction**

Probably the most important teleconnection phenomena are those of the El Niño–Southern Oscillation (ENSO) (Bjerknes, 1969; Neelin, 1998). The ENSO (Timmermann et al., 2018) is a natural, irregular fluctuation in the tropical Pacific region, and mostly affects the tropical and the subtropical regions; however, it has an impact on the global climate system as well. A crucial and open question that has challenged scientists for decades is how the ENSO would change as a result of the increasing radiative forcing due to the increasing greenhouse gas concentrations. The IPCC has low confidence in what would exactly happen to the ENSO in the future, even though they have high confidence that the ENSO itself would continue (Christensen et al., 2013). There have been several studies (e.g. Guilyardi et al., 2009; Collins et al., 2010; Vecchi and Wittenberg, 2010; Cai et al., 2015) that aimed to reveal how the ENSO might respond to greenhouse-gas forcing. However, most of the applied methods have a common drawback: they use temporal averages (including variances, correlations, etc.) in a time-dependent dynamical system, i.e., in our changing climate, or in simplified models thereof.

In a changing climate, where one or more relevant parameters are changing in time, there can be no stationarity by definition, whereas stationarity is crucial for the applicability of temporal averages, as illustrated by Drótos et al. (2015) in a toy model. In realistic GCMs globally averaged quantities seem to behave better, but the problem proves to be significant for local quantities and teleconnections (Herein et al., 2016, Herein et al., 2017). Since the ENSO events are identified by temperatures that are warmer or cooler than average, and teleconnections are defined as correlations between such anomalies, it is important to have a firmly established notion of averages when climatic means are shifting, as also pointed out by L'Heureux et al. (2013, 2017) and Lindsey et al. (2013).

To avoid the problem of evaluating time averages in a changing climate, in this study we turn to a gradually strengthening view according to which the relevant quantities of the climate system are the statistics taken over an ensemble of possible realizations evolved from various initial conditions (see e.g. Ghil et al., 2008; Bódai et al., 2011; Bódai and Tél, 2012; Deser et al., 2012; Daron and Stainforth, 2015; Kay et. al., 2015; Stevens, 2015; Bittner et al., 2016; Herein et al., 2016; Herein et al., 2017; Drótos et al., 2017; Hedemann et al., 2017; Lucarini et al., 2017; Suárez-Gutiérrez et al., 2017; Li et al., 2018). In contrast to weather forecast, one focuses here on long-term properties, independent of initial conditions, in order to characterize the internal variability, as well as the forced response of the climate. The mathematical concept that provides an appropriate framework is that of snapshot (Romeiras et al., 1990; Drótos et al., 2015) or pullback attractors (Arnold, 1998; Ghil et al., 2008; Chekroun et al., 2011); the concept's applicability has also been demonstrated in laboratory experiments (Vincze et al., 2017).

Qualitatively speaking, a snapshot attractor is a unique object in the phase space of dissipative systems with arbitrary, non-periodic forcing, to which an ensemble of trajectories converges within a basin of attraction. (In the climatic context, the ensemble members can be regarded as Earth systems evolving in parallel, all of which are controlled by the *same* physics and are subject to the *same* external forcing (Herein et al., 2017).) If the dynamics is chaotic, convergence implies that the initial condition of the ensemble is "forgotten": after some time (the convergence time), the evolution of the particular ensemble becomes independent of how it was initialized; instead, the distribution of its members, at any time instant, becomes determined by the natural probability distribution of the attractor. (This means that the ensemble members, in the given time instant, characterize the plethora of all possible weather situations permitted in the Earth system in a probabilistically correct way (Drótos et al., 2017).) The snapshot attractor and its natural distribution depend on time in general, and their time evolution is determined uniquely by the forcing scenario of the system. Note that ensemble statistics are instantaneous by construction. (Quantities characterizing a year or a month (e.g. a yearly average) can be considered instantaneous in a climate change of much longer time scale.)

In this paper we directly construct the snapshot attractor and its natural probability distribution, following Herein et al. (2017), and apply our new methodology to the relationship between the ENSO and the Indian summer monsoon. To our knowledge, it is the first time that the snapshot approach (taking care of the convergence) is used in the context of these phenomena. We analyze here the response of their relationship to scenarios with increasing radiative forcing in two different aspects: one is how the strength of the connection between the fluctuations of the two phenomena (called a 'teleconnection') responds, and the other one is a simultaneous observation of the response of climatic means characterizing each phenomenon.

## 2. Subjects of the study

Our investigations concern ensemble simulations from two state-of-the-art climate models: the Community Earth System Model (CESM, Hurrell et al. (2012)) and the Max Planck Institute Earth System Model (MPI-ESM, Giorgetta et al. (2013)).

The CMIP5 versions of these models were already studied regarding how reliable their ENSO characteristics are. It is known that both models underestimate the ENSO asymmetry, but all of the CMIP5 models suffer from this problem (Zhang and Sun, 2014). Generally, however, both models show relatively good ENSO characteristics compared to observations (Bellenger et al., 2014; Capotondi, 2013). The pattern of the monsoon precipitation is quite realistic in both models, however, the future projections for the Indian region generally have a moderate confidence (Freychet et al., 2015). In the recent study of Ramu et al. (2018), the strength of the ENSO–monsoon teleconnection has been found to be considerably underestimated in both models compared to observations. We must note, however, that Ramu et al. (2018) calculate the correlation coefficient with respect to time in a historically forced single run, so that the resulting values are possibly unreliable, cf. Section 6 and the Supplementary Material (part IV).

We consider four ensembles in total. The CESM community designed the CESM Large Ensemble ('CESM-LE') with the explicit goal of enabling assessment of climate change in the presence of internal climate variability (Kay et al., 2015). All realizations use a single model version (CESM with the Community Atmosphere Model, version 5) at a resolution of 192 × 288 in latitudinal and longitudinal directions, with 30 atmospheric levels. The MPI-ESM (Giorgetta et al., 2013) was also used to produce ensembles (called together the 'Grand Ensemble', 'MPI-GE') to explore internal variability in a changing climate (Stevens, 2015; Bittner et al., 2016). The single configuration applied for this purpose is model version MPI-ESM1.1 in low resolution (LR) mode, which corresponds to a horizontal resolution of T63 with 47 vertical levels in the atmosphere, and to 1.5-degree horizontal resolution with 40 vertical levels in the ocean.

The CESM Large Ensemble ('CESM-LE') consists of 35 comparable members, and covers the time span of 1920-2100. Between 1920 and 2005, historical climate forcing (Lamarque et al., 2010) is used, and the RCP8.5 (van Vuuren et al., 2011) is applied afterwards, reaching a nominal radiative forcing of $Q = 8.3$ W/m$^2$ by 2100. The MPI-ESM historical ensemble ('MPI-HE' in what follows) has 100 members and runs from 1850 to 2005 under historical climate forcing (Lamarque et al., 2010). The nominal radiative forcing becomes thus $Q = 2.1$ W/m$^2$ by 2005 (similarly as in the CESM-LE). The MPI-ESM RCP8.5 ensemble (which we shall call 'MPI-RCP8.5E') continues the previous runs between 2006 and 2099 under the RCP8.5 (van Vuuren et al., 2011) — by the time of our analysis, 77 members are available. Finally, the MPI-ESM 1-percent ensemble ('MPI-1pctE' in what follows) consists of 68 members, and, similarly to the MPI-HE, it starts in 1850, with the same (pre-industrial-like) external conditions. Being an idealized experiment, the $CO_2$ concentration is increased in this case by 1 percent per year until 1999, while the concentrations of other greenhouse gases and radiative agents are kept constant. The nominal radiative forcing (calculated via the logarithmic response (Ramaswamy et al., 2001)) reached by 1999 is $Q = 8.3$ W/m$^2$.

Fig. 1 gives an overview (Meinshausen et al., 2011) of the forcing scenarios, interpreted in terms of the nominal radiative forcing $Q$, in the timespans of our particular investigations (we discarded the beginning of each simulation except for the MPI-RCP8.5E — see later in this Section). Note that the nominal radiative forcing $Q$ is *not* a parameter of the system, so that its time dependence is *not* a forcing from a dynamical point of view. Instead, we treat it as a proxy for the aggregated effect of all different forcing agents (which include different tracers in the atmosphere, as well as the varying solar activity and land use — except for the MPI-1pctE).

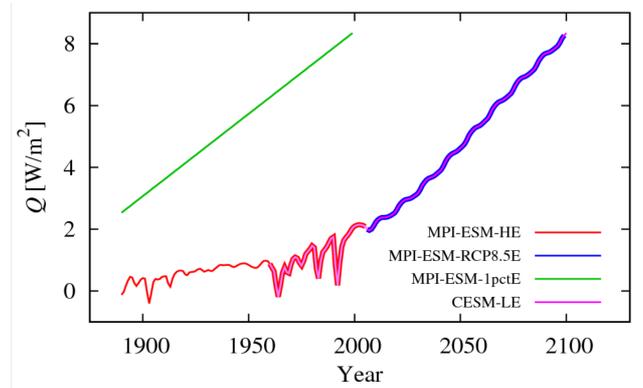

*Fig. 1. The nominal radiative forcing Q as a function of time in the particular simulations within the timespan of our investigation. For the nominal radiative forcing in the CESM-LE, in the MPI-HE and in the MPI-RCP8.5E, see Meinshausen et al. (2011). The nominal radiative forcing in the MPI-1pctE has been calculated via the logarithmic response (Ramaswamy et al., 2001).*

In order to ensure memory loss (i.e., convergence to the snapshot attractor (Drótos et al., 2015; Herein et al., 2016; Drótos et al., 2017)), in most cases we discard the first 40 years of the simulations (the only exception is the MPI-RCP8.5E). We emphasize that, in principle, a detailed and dedicated investigation should be carried out in both models to determine the time scale of the convergence, as advocated in Drótos et al. (2017). Due to technical limitations, however, this is far beyond the scope of the present study, which we believe to nevertheless provide with reliable results with the assumption of maximum 40 years for the convergence time, see Appendix A.

We note that our estimates for the convergence time correspond to the convergence properties that are determined by the atmosphere and the upper ocean, and *not* those that characterize the deep ocean and its abyssal circulation. Since the latter have time scales of thousands of years, it seems to be reasonable to assume that the members of the ensemble do not yet spread in the corresponding variables within the course of less than two hundred years (i.e., within the time span of our study). According to this time-scale separation, we conjecture that the adjustment of the slow climate variables corresponding to the abyssal circulation does not influence substantially the statistical properties investigated here. Note that, otherwise, all the studies on the 21st-century climate change performed by looking at the properties of an ensemble of simulations would be hopeless. The details of this time-scale separation in the climate system and its particular implications remain the topic of future research.

## 3. Characterizing the ENSO in a changing climate

The phases of the ENSO are traditionally characterized by looking at carefully constructed climate indexes, which surrogate the dominant features of the behavior of the climatic fields of interest. As a basis for the characterization, we consider the difference, denoted by $p_{\text{diff}}$, between the monthly or seasonal mean of the sea level pressure at Tahiti and at Darwin (see Appendix B). We intentionally choose such a simple variable, since it lacks the need of statistical preprocessing, thus avoiding possible corresponding ambiguities. The difference $p_{\text{diff}}$ is the basis of the Southern Oscillation Index (SOI) as defined by the Bureau of Meteorology of the Australian Government, also called the Troup SOI: an anomalously low (high) value of $p_{\text{diff}}$ indicates an El Niño (La Niña) phase (Troup, 1965).

In the Supplementary Material (part I), we recall from Herein et al. (2017) that climate indices should be treated carefully in a changing climate. In particular, long-term temporal averaging has to be avoided in their definition, and should be replaced by averaging with respect to the ensemble (after convergence has occurred). Indices or any anomalies defined in this proper way do not carry information about temporal shifts in the climatic mean of the corresponding quantities (like $p_{\text{diff}}$). Therefore, investigations of shifts in climatic means have to be carried out separately from those targeting the internal variability, an aspect of which is, however, characterized by anomalies or climate indices. In what follows, we shall present examples for both kinds of investigations.

**4. A forced response of the internal variability: the example of an ENSO teleconnection in a changing climate**

A special aspect of internal variability is the presence of teleconnections: for certain variables characterizing geographically distant regions, anomalies with respect to their climatic mean do not occur independently in a statistical sense. As an example in the case of the ENSO, if $p_{diff}$ is anomalously high during the summer months, there is a good chance that the precipitation of the Indian monsoon is also anomalously high.

The simplest way to quantify the strength of the (tele-) connection between two given variables is via Pearson's correlation coefficient $r$ (Rogers and Nicewander, 1988). Note that the correlation coefficient is obtained, by definition, as the average of the product of the anomalies (as defined by subtracting the average and dividing by the standard deviation) of the corresponding quantities. Consequently, a correlation coefficient between anomalies is the same as that between the original quantities. We underline that Pearson's correlation coefficient is limited to detect a linear correlation between the two quantities of interest; nonetheless, it is definitely useful for having a first order picture of the existing correlations in the fields.

In Herein et al. (2017) it has been demonstrated that the traditional evaluation of correlation coefficients, carried out via averaging over time, provides with grossly incorrect results. It is thus important to evaluate correlation coefficients with respect to the ensemble. As this can be done at *any* instant of time (after convergence), it also enables one to monitor the temporal evolution of the strength of the teleconnection during a climate change. This temporal evolution is one aspect of the response of internal variability to an external forcing. This is what we shall investigate in this Section for the teleconnection between the ENSO and the Indian monsoon.

In particular, we numerically evaluate the ensemble-based correlation coefficient using the sea level pressure difference between Tahiti and Darwin ($p_{diff}$), and a particular characteristic of the Indian monsoon, the precipitation over Northern India ($P$) (see Appendix B):

$$r = \frac{\langle p_{diff} P \rangle - \langle p_{diff} \rangle \langle P \rangle}{\sqrt{(\langle p_{diff}^2 \rangle - \langle p_{diff} \rangle^2)(\langle P^2 \rangle - \langle P \rangle^2)}} \quad , \tag{1}$$

where <...> denotes averaging with respect to the ensemble.

Although the temporal character of the forcing is quite different in certain ensembles, the results are more easily compared if we plot them as a function of the radiative forcing $Q$ instead of time. One should keep in mind, however, that the response is always expected to exhibit some delay (Herein et al., 2016), and that the nominal radiative forcing $Q$ is just a proxy for the aggregated effects of different forcing agents (see Section 2). The results are presented in Fig. 2 for all ensembles considered. Due to the moderate size of the ensembles, especially for the CESM-LE but also strongly affecting the MPI-ESM ensembles, the numerical fluctuation of the signals are considerable, so much that one cannot read off meaningful coefficients for particular years (corresponding to individual data points in our representation). The fine structure of the time dependence thus remains hidden. What might be identified, however, from our plots are main trends (or their absence), with approximate values on a coarse-grained temporal resolution. Had our ensembles been of infinite size and, thus, able to describe accurately the distribution supported by the snapshot attractor, we would be able to have information at all time scales.

We focus in Fig. 2 on the Indian summer, since we did not find considerable correlations for the spring and the autumn (as their precipitation is not of monsoonal origin), and the bounded nature of the precipitation (from below) introduces a strong nonlinearity for the winter. In Fig. 2a, the whole JJA season is taken, the main period of the Indian summer monsoon. Here the MPI-ESM ensembles seem to give a constant value, ≈0.4, for the coefficient, both when plotted as a function of $Q$ and when plotted as a function of the time $t$ (see the inset). By a visual inspection, no trends can be identified even for the MPI-RCP8.5E or the MPI-1pctE. The value itself of the correlation coefficient is in harmony with the observations (Walker and Bliss, 1937; Parthasarathy and Pant, 1985). At the same time, the CESM shows very little positive correlation for JJA (with no trend either). Such a large discrepancy is unexpected. Nevertheless, the underestimation of the strength of the teleconnection by the CESM agrees with Ramu et al. (2018). The absence of such an agreement with Ramu et al. (2018) for the MPI-ESM may stem from the unreliability of correlation coefficients evaluated with respect to time (see Section 6), but also from characterizing the ENSO by a different variable.

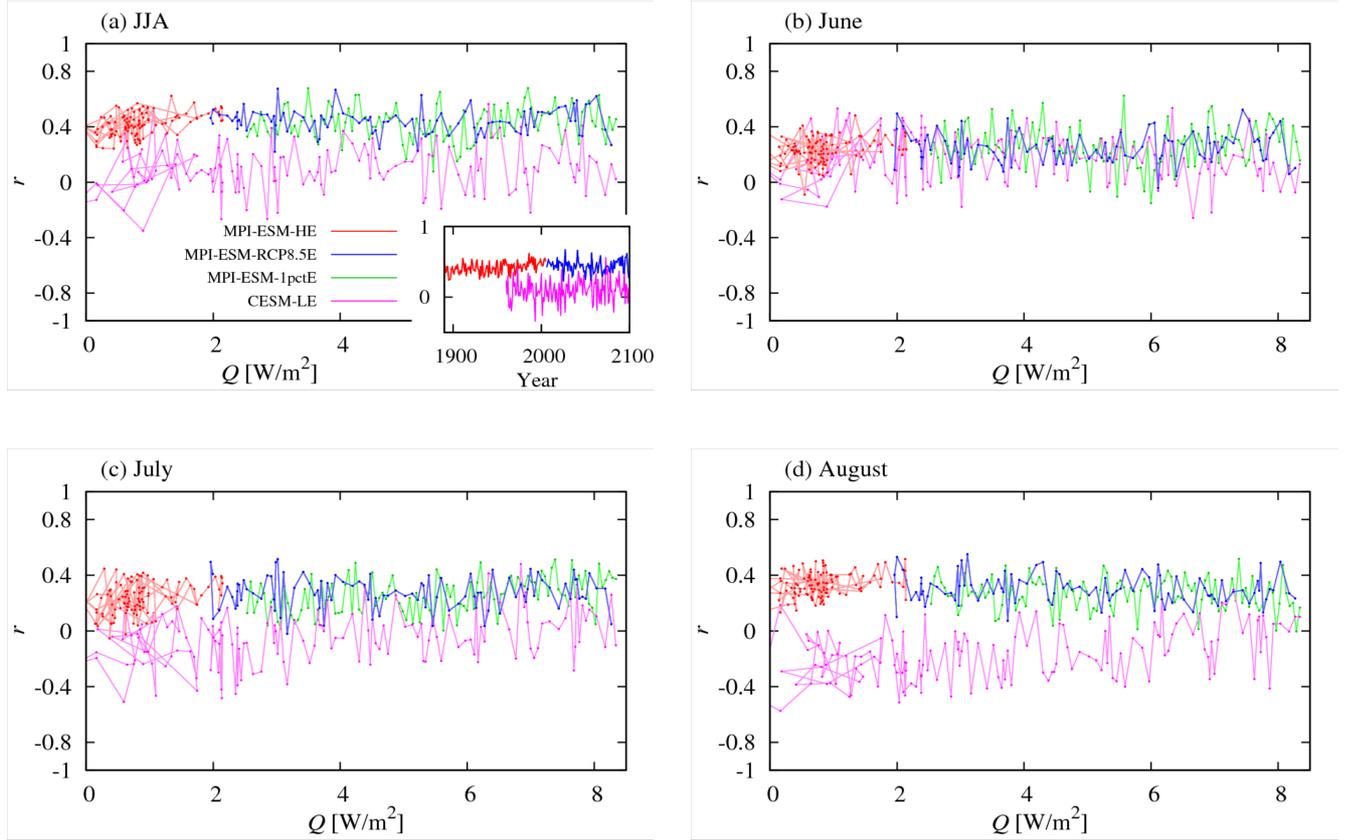

*Fig. 2. The correlation coefficient r, plotted as a function of the radiative forcing Q, between the sea level pressure difference $p_{diff}$ and the Northern Indian precipitation P, in all three ensembles. The consecutive years are connected by lines. In the different panels, different intervals of the year, as indicated, are considered for which $p_{diff}$ and P are averaged. In the inset of panel (a), the time evolution is also given as a function of the time t (omitting the MPI-1pctE for better visibility).*

In the further panels of Fig. 2, the characteristics of the individual months are considered. Fig. 2b shows that the two models approximately agree for June: all ensembles give a constant, positive coefficient, with a value of about 0.2 or 0.3. The coefficient of the CESM-LE is again slightly lower than those of the MPI-ESM ensembles, but the difference is not large. However, for July and August (Figs. 2c and 2d, respectively) the CESM-LE gives a much lower correlation coefficient than the MPI-ESM ensembles. In fact, the former coefficient is undoubtedly *negative* in August, which is in contrast with the observations and the general expectations. This different model behavior (of unknown origin) in July and August is what underlies the considerably smaller coefficient for the whole JJA season.

Note also in Fig. 2 that the correlation coefficient is typically smaller for any individual month than for the whole season, in any given ensemble.

After the visual inspection, we take to formally testing if the seasonal correlation coefficient for JJA is constant during the timespan of each of the four simulations. Posing a null hypothesis of a changing coefficient leaves the problem underdetermined, hence we adapt the null hypothesis of a constant coefficient. Most of our tests are based on the fact that the distribution of the Fisher-transform (Fisher, 1915; 1921), i.e., the area hyperbolic tangent, which we shall denote by *z*, of an estimate of a given correlation coefficient *r* calculated from an ensemble of given size follows, to a very good approximation, a known distribution: a Gaussian with a standard deviation of $1/\sqrt{(N-3)}$, where *N* is the ensemble size (Fisher, 1936).

From our null hypothesis, it would follow that the Fisher-transforms *z* of the estimates of our hypothetical constant coefficient calculated in the different years would be independently drawn samples from this known distribution. (For this exceptional, single variable, independence is supported by the calculations described in part II of the Supplementary Material.) Thanks to the Fisher-transform resulting in a known distribution, we first test by a Kolmogorov-Smirnov test if the estimates from the entire time series (denoting the corresponding p-value by $p_{KS0}$), and those from the first ($p_{KS1}$) and the second half ($p_{KS2}$) of the time series may originate from that known distribution with arbitrary mean. Second, assuming "yes" for an answer, and given that the known distribution resulting from the Fisher-transform is the normal distribution, we can test by a *t*-test if the estimates in the first and the second half of the time series may come from populations with the same mean ($p_{t12}$). Note that we split the data in to two halves with respect to *time*, so that we are testing for the stationarity of the time series in the usual, temporal sense. Third, to complement the *t*-test, we test for the presence of a monotonic trend (as a function of time) by a Mann-Kendall test, without any assumption for the shape of the distribution ($p_{MK0}$). See Appendix C for the description of the particular tests.

The numerical results in Table 1 for $p_{KS0}$, $p_{KS1}$ and $p_{KS2}$ indicate that the time series of all simulations, either in part or in total, may originate from the hypothesized (stationary) distribution, since all of these p-values are larger than any usual significance level. Surprisingly, we cannot reject stationarity for any of the MPI-RCP8.5E, the MPI-1pctE, and the CESM-LE according to $p_{t12}$ and $p_{MK0} > 0.05$; meanwhile, $p_{t12}$ and $p_{MK0}$ being much smaller than the usual significance levels, stationarity can be rejected for the MPI-HE. This is rather interesting, because the MPI-HE is subject to the weakest forcing within the considered set of simulations. In the Supplementary Material (part II) we present specific, additional arguments that the very low $p_{MK0}$ value in the MPI-HE is not corrupted measurably by possible autocorrelations in the time series.

One might think that the ability to pose stronger statements for the MPI-HE originates from its larger ensemble size. In the Supplementary Material (part III) we illustrate that it may only be the case for the CESM-LE, but not for the other ensembles.

|  | $p_{KS0}$ | $p_{KS1}$ | $p_{KS2}$ | $p_{t12}$ | $p_{MK0}$ |
|---|---|---|---|---|---|
| MPI-HE | 0.99 | 0.93 | 0.92 | 0.00020 | 0.000021 |
| MPI-RCP8.5E | 0.75 | 0.28 | 0.93 | 0.76 | 0.58 |
| MPI-1pctE | 0.60 | 0.33 | 0.42 | 0.64 | 0.25 |
| CESM-LE | 0.83 | 0.68 | 0.99 | 0.072 | 0.20 |

*Table 1. The* p-*values of several hypothesis tests in the different ensembles. See text for details.*

We also conclude that the correlation coefficient may only *increase* from the first to the second half of the simulation in the case of the MPI-HE, since $p_{KS1} < 0.05$ for all those possible mean values of the above-mentioned distribution for which $p_{KS2} > 0.05$, and vice versa. For a complete view, see Fig. 3.

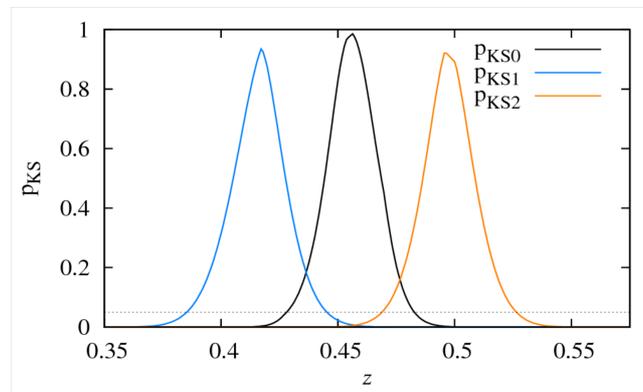

*Fig. 3. The* $p_{KS0}$, $p_{KS1}$ *and* $p_{KS2}$ *values as a function of the mean of the Gaussian distribution from which the data is tested to originate.*

By scanning through all possible subintervals of the time span of the MPI-HE, and calculating the p-value of the Mann-Kendall test ($p_{MK}$) for each, we find that the smallest $p_{MK}$ values (those that are significant on the level of 0.05) occur mainly in intervals with endpoints from the last two decades of the simulation (see Fig. 4). That is, the increase occurs mostly in these decades.

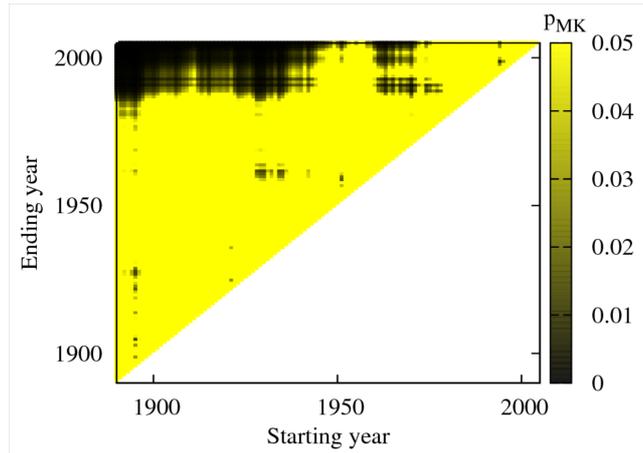

*Fig. 4. The $p_{MK}$ values (color coded) calculated in the MPI-HE for all possible subintervals of its time span. Yellow corresponds to $p_{MK} = 0.05$ or larger.*

The results presented in this Section are impossible to obtain with traditional techniques (which evaluate correlation coefficients over time in single realizations), as illustrated in the Supplementary Material (part IV).

It is worth trying to make use of the fact that response is detectable only in the radiatively most weakly forced setup (i.e., in the MPI-HE) to draw some conclusion about the dynamical role of the radiative forcing $Q$. In the Supplementary Material (part V), we conclude that either the response in the strength of the teleconnection with respect to $Q$ is very different from linear, or $Q$ cannot be regarded as dynamical forcing from this point of view.

**5. The forced response of the climatic mean of two quantities**

Certain aspects of the forced response of the climate system can be represented by the displacement of the climatic mean in projections of the phase space. A point marking the instantaneous climatic mean in such a space represents a statistical signature of *all* kinds of weather situations permitted in the given time instant. If two such points in this space do not coincide, it means that the dynamical structure (the attractor itself), or, equally, the plethora of all possible weather situations, is different for the given two cases. In the language of thermodynamics, such a climatic mean is thus a "state indicator", characterizing the "macroscopic" state of the system, but not its "microscopic" state (i.e., the particular realization). Plotting a point for each time instant during a climate change enables one to follow the evolution of this "state indicator" (characteristic of the dynamical structure, i.e., the attractor), clearly separated from the temporal fluctuations, which appear in any particular realization of the system, and which represent the internal variability.

This analysis of the forced response of the climatic mean is already meaningful for one variable, as shown in Supplementary Fig. S2, and also in Drótos et al. (2015) and Herein et al. (2016). Actually, the term "global warming" (one specific aspect of climate change) refers to the increase in the climatic mean of the global surface temperature. If we use a phase space of more than one quantities, we can carry out an even more informative investigation: we can study if we find special relationships between the climatic means of these quantities. What we consider particularly interesting is to choose quantities whose fluctuations are already known to be correlated ("teleconnected") so that we might expect their climatic means to respond to external forcings in a coordinated way as well. One possibility could be a universal functional relationship, which could reflect some robust mechanism, not affected by climate change, linking the investigated quantities. In the following, we focus on the pressure difference $p_{diff}$, which is related to the ENSO, and on the Northern Indian precipitation $P$, which is a characteristic of the Indian monsoon. To our current study, we shall include the investigation of the seasonality.

Since no considerable climatic shifts are observable in the MPI-HE, and the MPI-1pctE is extremely similar to the MPI-RCP8.5E from the current point of view, we show, in Fig. 5, numerical results for the two other ensembles only. (For the MPI-1pctE, see Supplementary Fig. S6.) The seasonal cycle is clear and has a very similar shape in both models, with the Northern Indian precipitation P concentrating on the months June-October, and the pressure difference pdiff increasing from June to December, with a "stalling" from September to October. Although the magnitude of the precipitation is similar in the two models, that of the pressure difference is very different: the CESM exhibits a 200hPa offset to smaller values.

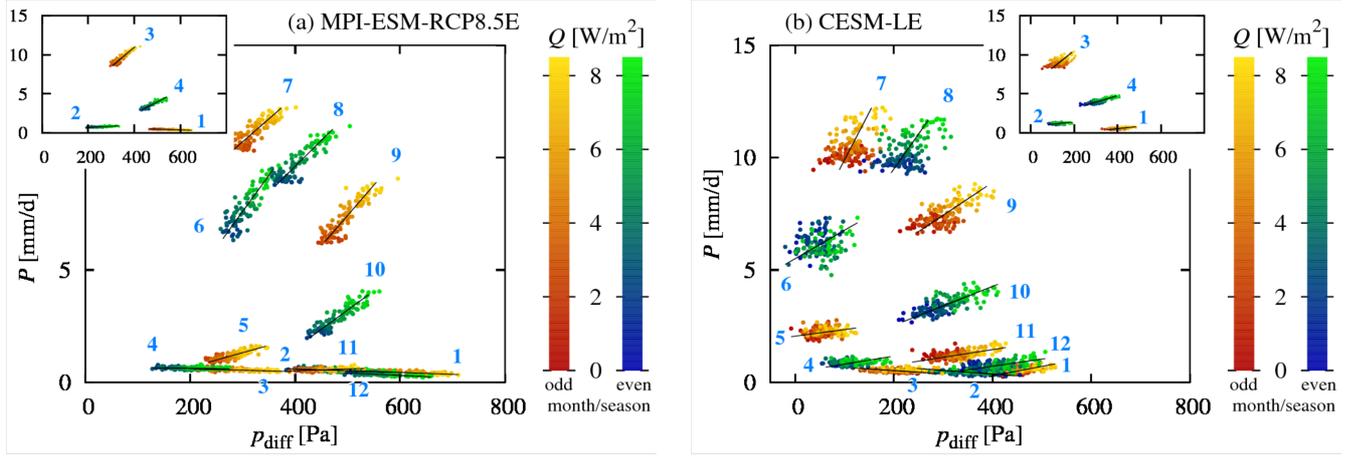

*Fig. 5. The climatic mean (obtained as the ensemble average) in the sea level pressure difference $p_{diff}$ and the Northern Indian precipitation P. All different months (in the main plots) and seasons (in the insets) are plotted, see the numbering (1-12: January-December, 1-4: DJF-SON). For a given month or season, each data point represents a particular year. The different years are colored according to the color scales on the right. The thin black lines correspond to the slopes fitted by weighted total least squares (Krystek and Anton, 2007). (a) MPI-RCP8.5E, (b) CESM-LE.*

In both ensembles, the response of the climatic mean closely follows a linear shape in the phase space projection of Fig. 5: both quantities $p_{diff}$ and $P$ increase. The increase of these quantities agrees with the literature (Power and Kociuba, 2011; Kitoh et al., 2013; Li et al., 2015; Pascale et al., 2016). The linear shape might be a consequence of staying in the regime where the system's response to the forcing is linear (Lucarini et al., 2017). The slopes are determined by the susceptibilities of the quantities in question to the applied forcing, and we numerically estimate these slopes by fitting lines with the method of the weighted total least squares (Krystek and Anton, 2007). The error of a given data point in each variable is assumed to be proportional to the standard deviation (taken with respect to the natural distribution) in that given variable in the particular year.

Note that in Fig. 5 the slopes of the fitted lines are different in the different months or seasons. As a consequence, there can be no universal functional relationship entangling the climatic means of the variables $p_{diff}$ and $P$, without involving additional variables, that would be valid at all time scales. The fact that the slope differs from month to month means that the linear response differs when considering different slices of the year. That is, not just the climatic means of $p_{diff}$ and $P$ fail to satisfy a universal relation, but this implies that nor can the response of these means to a given forcing be related by some function linking only these two variables.

Such a relationship between the response of the snow cover and that of the surface albedo was found by Qu and Hall (2007) by comparing different models. Similar, so-called emergent relationships between responses of different quantities, and also between different types of responses of a single quantity, have since been found in model ensembles, see e.g. Cox et al. (2013), Wenzel et al. (2016), Cox et al. (2018). As a common feature, they are hoped to have the potential to reduce the uncertainty in climate projections, i.e., to form the basis of so-called emergent constraints. Our example illustrates that comparing different seasons instead of different models can also be a way of looking for emergent constraints. One should note, however, that even if a season-based study rejects a universally present emergent relationship between certain variables, like here, it does not exclude the presence of an emergent relationship in a model ensemble if one given slice of the year is compared.

The strong seasonality in our case may be due to the seasonally varying circulation patterns that dominate the investigated regions. Note that the corresponding teleconnection, studied in Section 4, is observed only for the summer Indian precipitation, which means that it is also strongly affected by seasonality.

The most prominent feature of the seasonality of the response of the climatic means (Fig. 5) is as follows. The precipitation $P$ responds strongly only in and after the months of the Indian monsoon (June-October), in the form of a large increase, while the other months remain mostly dry. At the same time, the pressure difference $p_{diff}$ does not exhibit an enhanced increase in June-September compared to other parts of the year. In the CESM-LE, actually, $p_{diff}$ exhibits almost no increase in June and July (see the almost vertical lines in Fig. 5b); the increase in $p_{diff}$ is only considerable in the later part of the year, from September, during which the increase in the precipitation $P$ gets weaker (as indicated by the lines turning to nearly horizontal from August to November in Fig. 5b). Such an absence of a summer response in $p_{diff}$ is not observed in the MPI-RCP8.5E (Fig. 5a). Nevertheless, both models show that one variable can undergo strong changes without corresponding in-phase changes in the other variable.

We note that other choices for the representation of the East-West pressure difference may lead to different trends, cf. Power and Kociuba (2011) and also Vecchi et al. (2006). The phenomenology presented here, however, is not expected to be specific to our choice.

We add that we found it impossible to approximate the correct results to some useful degree when we used the traditional temporal averaging to obtain the climatic means instead of averaging over the ensemble. In particular, we encountered very strong false trends. See the Supplementary Material, part VIII, for details.

## 6. Discussion

In this paper we analyzed ensemble simulations in the CESM and the MPI-ESM subject to historical forcing, to a 1 percent per year increase in the $CO_2$ concentration, and to the RCP8.5. We were interested in the relationship between simple variables characterizing the ENSO and the Indian summer monsoon, both in terms of the fluctuations (the teleconnection) and in terms of the joint response of the climatic means of these variables.

We found that the MPI-ESM satisfactorily reproduces the expected features of the teleconnection, while the CESM behaves more unexpectedly. In the MPI-ESM, the teleconnection undergoes a considerable strengthening in the modeled historical period between 1890 and 2005, concentrated to the last decades of the simulation. In the same model, however, change is detected neither between 2006 and 2099 under the RCP8.5, nor in a time interval of 110 years under a 1-percent pure $CO_2$ scenario — in both of the latter scenarios, the radiative forcing is practically always higher, and covers a much wider range than in the historical forcing scenario (see Fig. 1). Furthermore, change is not detected in the 1960-2100 time interval simulated according to historical forcing and the RCP8.5 in the CESM either.

From an analysis of the sensitivity of our hypothesis tests of stationarity (Supplementary Material, part V), we conclude that the strength of the teleconnection cannot respond to radiative forcing instantaneously and linearly, since our tests would have had to detect nonstationarity in the ensembles other than MPI-HE, too. Lags of few years are not expected to change the picture much, but some very strong form of nonlinearity could explain the results in principle. However, a more plausible explanation, in our opinion, is that assuming the radiative forcing $Q$ to be dynamical forcing (i.e., a single quantity that appears explicitly in the equations of motion) is not a good approximation.

In particular, the strength of the teleconnection may respond in a different way to variations in different forcing agents. Remember that the nominal radiative forcing $Q$ represents the aggregated effects from several different agents. It is even possible that the increase in the nominal radiative forcing $Q$ and in the strength of the teleconnection is just a coincidence, the latter possibly resulting from mechanisms not or not directly related to the increase in the net energy flux. In any case, responses might not be possible to be interpreted in terms of variations in the single quantity $Q$.

In fact, the differentiation of responses with respect to different forcing agents would not be very surprising. Considering a (globally homogeneous) $CO_2$ forcing alone, while the resulting $Q$ might act as a dynamical forcing with respect to the surface temperature, it might not do so with respect to other observables, such as the temperature at the tropopause (Bodai et al., 2018). The teleconnection of the ENSO with the Indian summer monsoon might indeed involve a physical mechanism not restricted to the surface, or to observables that would secure the causality of the "response" of the teleconnection to $Q$. If we add that the response in the climatic mean of the Indian summer monsoon has actually been found by Li et al. (2015) to be governed by different mechanisms under aerosol forcing (related to volcanism) and greenhouse-gas forcing (although utilizing techniques based on temporal averaging, see later in this Section), we can easily imagine that the fluctuations of the Indian summer monsoon respond differently to these two kinds of forcings, causing the teleconnection respond in a different way, too.

Note that volcanism is enhanced in the late 20$^{th}$ century when changes in the strength of the teleconnection are most prominent. In fact, a hypothesis has been put forward by Maraun and Kurths (2005) that after major volcano eruptions in the Southwest Pacific the "cooling effect could reduce the land/sea temperature gradient and thus make the Monsoon more sensitive to ENSO influence". These authors found more regular oscillatory ENSO dynamics and a phase locking between the ENSO and the monsoon in the observed time series after major volcano eruptions in southern Indonesia, which should be reflected in an increased correlation, consistent with our finding. This could also be an indication that in terms of a nonlinear quantifier of the teleconnection a single realization contains already a lot of information about the forced response — the latter of which we always define in terms of an ensemble.

The increasing strength of the studied teleconnection in the last decades of the 20$^{th}$ century in the MPI-ESM is in a complicated relationship with the consensus about its *decreasing* strength in the late 20$^{th}$ century (Krishna Kumar et al., 1999; Kinter et al., 2002; Sarkar et al. 2004; Maraun and Kurths, 2005; Boschat et al., 2012). Note that there is no consensus about the evolution of this strength in 21$^{st}$-century climate projections (Ashrit et al., 2001; Ashrit et al., 2003; Ashrit et al., 2005; Annamalai et al., 2007; Li and Ting, 2015; Ruiqing et al., 2015).

As for the 20th-century behavior, Krishna Kumar et al. (1999) suggest that the decrease has a statistical significance, and so it should be a signal of forced response, for which global or Eurasian continental warming is responsible (the reverse of the effect of volcanism enhancing the connection). Maraun and Kurths (2005) also argue that the decrease is a forced response, but claim that, beside global warming, it should also be due to volcanism. They describe a transition near 1980 from 1:1 phase locking into a 2:1 phase locking, with the Indian monsoon oscillating twice as fast. This connection would be "invisible to (linear) correlation analysis", or rather the correlation would be destructed by the additional monsoon peak. Clearly, the result of the MPI-HE would be opposing these conclusions. However, evaluating correlation coefficients with respect to time, as e.g. in Krishna Kumar et al. (1999), especially if done without some detrending, is methodologically incorrect, as we shall discuss later in this Section. This means that the MPI-HE, to our knowledge, does *not* disagree with observations about the presence or the sign of forced trends.

Actually, it is recognized in many studies (Kinter et al., 2002; Ashrit et al., 2003; Ashrit et al., 2005; Annamalai et al., 2007; Kitoh, 2007; Chowdary et al., 2012; Li and Ting, 2015) that "modulations" and corresponding trends in the studied correlation coefficient, when calculated over different time intervals (as done by e.g. Boschat et al. (2012) and Ruiqing et al. (2015)) or over moving (sliding) time windows (as done by e.g. Krishna Kumar et al. (1999), Ashrit et al. (2001), Kinter et al. (2002), Ashrit et al. (2003), Ashrit et al. (2005), Annamalai et al. (2007), Kitoh (2007), Chowdary et al. (2012), Li and Ting (2015)), can appear as a result of internal variability. In particular, Li and Ting (2015) conclude that the observed weakening of the teleconnection in the late 20th century would be due to internal variability. Since our finding in the MPI-HE concerns a response to forcing, it should not be compared to trends that result from internal variability.

We argue that meaningfully defining such trends is not even straightforward. As a principal interpretation, a correlation coefficient characterizes the co-occurrences of anomalies over all possibilities permitted by the underlying probability distribution (by the internal variability itself), and any "modulations" can only result from an incorrect sampling of this probability distribution (which is the natural distribution supported by the attractor of the climate system). With a correct sampling, a change, "modulation", in the correlation coefficient is *not* possible in the absence of some time-dependent external forcing, which would alter this distribution.

Defining "trends due to internal variability" is, however, possible in the sense discussed as follows. One should first note that low-frequency components of internal variability can indeed lead to longer periods of time when the relationship between the anomalies of the investigated variables appears to be stronger or weaker, exhibiting trends between such periods. Recognizing such periods and trends would undoubtedly be of practical relevance. From a technical point of view, such correlations can be described in terms of a *conditional* probability distribution, which is obtained by fixing the value of the variable or variables that describe the given low-frequency mode. That is, the full range of internal variability is *not* taken into account in this case. Note that earlier studies (Kinter et al., 2002; Ashrit et al., 2003; Ashrit et al., 2005; Annamalai et al., 2007; Kitoh, 2007; Chowdary et al., 2012; Li and Ting, 2015) do not specify what the low-frequency modes are, or which variables are considered to be fixed, so that their notion of correlation coefficients cannot be considered complete.

It is important to recall here that the technique of moving (sliding) time windows artificially introduces dominating low-frequency characteristics to the observed signals, as pointed out in general by Wunsch (1999), and specifically for the ENSO-monsoon teleconnection by Gershunov et al. (2000) and Yun and Timmermann (2018) (see part IV of our Supplementary Material, too). It is thus hard to judge if the trends and "modulations" reported in the literature are such artifacts or correspond to true low-frequency modes of internal variability. The mentioned effect should also be kept in mind, of course, when looking for forced trends in the "true", unconditional correlation coefficient (that fully characterizes internal variability). (Even taking an ensemble mean of such correlation coefficients calculated in individual members of the ensemble would not eliminate the spurious properties of the resulting signal unless the ensemble size were asymptotically large.)

Beyond characteristics introduced by the application of moving windows, two pitfalls should be discussed that concern correlation coefficients that are evaluated with respect to time, even over some fixed time window. The first is that some possible temporal autocorrelation in the investigated variables effectively reduces the size of the sample, which has to be taken into account e.g. when calculating significance levels and confidence intervals. The second applies in the presence of time-dependent forcing: in this case, different time instants are characterized by different probability distributions (cf. Section 1), so that any statistical quantifier loses its probabilistic meaning in a strict sense. Even in cases when the structure of internal variability changes only little, forced trends in the investigated variables can ruin the meaningful calculation of correlation coefficients, as demonstrated in Herein et al. (2017) and in part I of the Supplementary Material. A reasonable evaluation can only be carried out after removing such trends. Previous works in the literature (see above) usually do not discuss if a detrending was carried out for the analysis (exceptions are e.g. Ashrit et al. (2003), Boschat et al. (2012) and Chowdary et al. (2012)). Although we found detrending to be unimportant in the MPI-HE (see Supplementary Material, part IV), stronger trends present in the climatic means in 21st-century-like simulations (see Section 5) indicate the importance of this issue.

Taking into account all considerations discussed above, dedicated further investigations are required to decide if the increasing teleconnection strength found in the MPI-HE is in actual contradiction with instrumental records or not, and, separately, finding relevant low-frequency modes in the MPI-HE or other ensembles may also be interesting.

As a complementary investigation to that of teleconnections, visualizing the forced response of the climatic mean in low-dimensional projections of the phase space reveals separate information about relations or their absence between the selected quantities. Unlike for the teleconnection, we found obviously strong displacements in the phase space projection spanned by the Tahiti–Darwin sea-level pressure difference and the Northern Indian precipitation. In spite of these strong displacements, a linear relation between the response of these two quantities was revealed, which potentially originates from staying in the linear response regime. However, the slope exhibits a very strong seasonality, falsifying a hypothesis of a universal, emergent functional relationship (possibly forming the basis of an emergent constraint (Qu and Hall, 2007)) between these two quantities without involving further variables. As for the Northern Indian precipitation alone, we have shown that the summer monsoon gets much stronger in scenarios with an increasing radiative forcing.

Depending on the mechanism that gives rise to the emergentrelationship, our finding might have the important practical implication that the lack or presence of such a relationship can be prompted by a seasonal analysis of a *single* model instead of comparing different models. If seasonality has an effect on the relationship between certain variables, like in our case, any emergent relationship that possibly exists between the investigated variables must involve further variables, too, and this has to be taken into account when looking for possibilities of formulating emergent constraints.

One should note the contrast that the strength of the teleconnection seems to change more in the historical period, while the climatic means are displaced more in the 21st-century-like simulations. This underlines the importance of separating the investigation of fluctuations (internal variability) from that of climatic means.

An additional important experience, illustrated in the Supplementary Material (part VIII), is that temporal averages evaluated with running windows over single time series can produce strongly misleading trends for climatic means in GCMs (cf. Drótos et al., 2015). This is again a manifestation of the effects discussed in Wunsch (1999). Note that this is more than an issue of interpretation, since these characteristics are absent in the signals of the true climatic means (taken with respect to the natural probability distribution). This observation provides another warning, beyond what is discussed earlier in the Section in relation with correlation coefficients, about evaluating statistics with respect to time.

In general, the error of temporal averages (Drótos et al., 2016) or other statistical quantifiers evaluated over the time evolution of a single realization needs to be investigated separately in different circumstances, and their application in systems with time-dependent forcing especially carries a risk of obtaining misleading or incorrect results. (This includes more sophisticated statistical techniques, like the evaluation of empirical orthogonal functions, or comparing such constructions to data from nonstationary systems even if they are obtained from stationary counterparts of these systems, like in Li and Ting (2015) and Li et al. (2015) in our context.) Our results clearly illustrate that the climate indices that are traditionally used to characterize the ENSO (or any other climate feature) have to be updated and substituted by snapshot-based indices if the forcing is time-dependent (see part I of the Supplementary Material).

We underline that the snapshot framework provides with the conceptually correct probability distribution for all variables of the climate system, it is mathematically established and is devoid of any subjectivity, and it can separate externally induced trends from internal fluctuations (including the characterization of teleconnections). We therefore call the attention of the climate research community to the importance of evaluating any statistics in long-term climate projections within this framework, preferably with increased ensemble sizes.

**Acknowledgement**

The authors wish to express gratitude to J. Broecker, C. Franzke, T. Haszpra, T. Kuna, N. Maher, S. Milinski, and T. Tél for useful remarks and discussions. TB would like to thank F. Nijsse for discussions about emergent constraints. GD is thankful to B. Stevens, T. Mauritsen, Y. Takano, and N. Maher for providing access to the output of the MPI-ESM ensembles. The authors also wish to thank the Climate Data Gateway at NCAR for providing access to the output of the CESM-LE.

**Funding**

The simulations for the MPI-RCP8.5E were supported by the H2020 grant for the CRESCENDO project (grant no. 641816). MH is thankful for the great support of the of the DFG Cluster of Excellence CliSAP. This research was supported by the National Research, Development and Innovation Office - NKFIH under grants PD124272, K125171 and FK124256. GD was supported by the Ministerio de Economía, Industria y Competitividad of the Spanish Government under grant LAOP CTM2015-66407-P (AEI/FEDER, EU). TB, VL and FL were supported by the H2020 grants for the CRESCENDO (grant no. 641816) and Blue-Action (grant no. 727852) projects. VL has been supported by the DFG Sfb/Transregio TRR181 project.

**Conflict of Interest**

The authors declare that they have no conflict of interest.

**Appendix A. Initialization and convergence**

The CESM-LE has initial conditions obtained as minor perturbations of a single trajectory, i.e., very localized in the phase space, it is thus obligatory to discard the beginning of the simulation, until the trajectories sample the natural probability distribution correctly. A safe estimate for this convergence is 40 years, which originates from the discussion of the relevant time scales in Kay et al. (2015) and from additional CESM results (Kim et al., 2017).

As for the MPI-ESM, we suppose, on the one hand, that the relevant time scales are similar in different climate models of similar complexity in similar setups, and, on the other hand, preliminary results indeed suggest that the duration of a primary convergence should be considerably shorter than 40 years. Furthermore, the initialization scheme is very favorable in the MPI-GE.

In particular, the initialization of each member in the MPI-HE and the MPI-1pctE is done by picking a particular time instant from a 2000-year-long pre-industrial control run, in which external conditions are kept at the 1850 level eternally. The different members thus sample the complete (atmospheric) attractor corresponding to 1850, but they might not be perfectly uncorrelated from each other, since the average time between their initializations is around 20 years, which is less than the 40-year "safety limit" for the convergence and for the corresponding memory loss. As the total length of the preindustrial run, 2000 years, is much longer than 40 years, the preindustrial attractor is nevertheless sampled correctly by the initial conditions, and their potential correlation can be interpreted as a smaller effective size of the ensembles at the beginning of the simulations (up to e.g. 1890).

While this is one reason to skip this initial period for the computations, a more important reason is that the stationary attractor of the preindustrial control run, obtained by assuming conditions of 1850 eternally, is not identical to the snapshot attractor of 1850 that is determined by the historical forcing scenario (Lamarque et al., 2010) before 1850. Although this difference is supposed to be small, it follows from its presence that also afterwards, in the initial years of the simulation, the members of the MPI-HE are not yet distributed according to the natural distribution of the historical forcing scenario. This is so only after convergence takes place to the latter distribution (i.e., the 1850 initial conditions are forgotten). Our "safe estimate" is 1890 for the time when the convergence is ready, and we believe that this is a strong theoretical argument for discarding the preceding period from the MPI-HE (unlike in e.g. Hedemann et al. (2017)). For consistency and comparability, it seems to be reasonable to discard this period from the MPI-1pctE, too. However, as for the MPI-RCP8.5E, the initial conditions of its members are the endpoints of the trajectories of the MPI-HE, which have already converged by 1890, i.e., earlier than the beginning of the MPI-RCP8.5E in 2006. Therefore, discarding of initial years is not needed in the MPI-RCP8.5E.

**Appendix B. The numerical calculation of $p_{\text{diff}}$ and $P$**

For the sea-level pressure difference $p_{\text{diff}}$ between Tahiti and Darwin, we took the sea-level pressure at the models' gridpoints that are closest to (17º31' S, 21º26' E) for Tahiti and (12º28' S, 130º50' E) for Darwin, and calculated their difference.

For the precipitation over Northern India, $P$, a rectangular box domain has been used with its corners located at the following coordinates: (31º N, 76º E), (31º N, 88º E), (17º N, 76º E), and (17º N, 88º E). We chose this box as it contains the observed monsoon precipitation maxima along the Indian and the Nepalese Himalayas, which also appear in the models. The precipitation $P$ is obtained as the spatial mean of the total precipitation over this rectangular domain.

**Appendix C. Hypothesis tests**

For $p_{KS0}$, $p_{KS1}$ and $p_{KS2}$, we use the Kolmogorov-Smirnov test that tests whether some observed values may originate from a Gaussian distribution with a given mean and standard deviation (Massey, 1951). While the latter is determined by our null hypothesis, the mean is unknown. In the lack of any a priory assumption for the mean, we sweep through a very wide neighborhood of the mean value which we estimated from the observed data, with high resolution, and select the highest p-value that we encounter. We carried out the said Kolmogorov-Smirnov tests using the Matlab function 'kstest'.

The value of $p_{t12}$ is calculated from the unpaired two-sample *t*-test (Fisher, 1936), which assumes that the two observed datasets come from Gaussian distributions with the same standard deviation, and tests whether these two Gaussian distributions may have the same mean. In our case, $p_{KS1}$ and $p_{KS2}$ is intended precisely to test for deviations from the mentioned assumptions in the two datasets to be compared. The results for $p_{KS1}$ and $p_{KS2}$ (see Section 4) does not exclude that the assumptions are fulfilled, so that the deviation from them cannot be extremely large. If we additionally consider that the *t*-test is relatively robust against deviations from its assumptions (Markowski and Markowski, 1990; Lumley et al., 2002), we conclude that carrying out the *t*-test provides with meaningful results. We carried out the said t-tests using the Matlab function 'ttest2'.

The Mann-Kendall test (Mann, 1945; Kendall, 1975), which we use for $p_{MK0}$, tests the null hypothesis of the absence of a monotonic trend, without any assumption for the shape of the distribution. It is thus ideally suited for our purpose of testing stationarity if we suppose that a relatively weak forcing, like those in the ensembles studied in this paper, cannot result in nonmonotonic trends. Furthermore, if the absence of a monotonic trend can be rejected, this implies that stationarity can also be rejected. We carried out the Mann-Kendall and the modified Mann-Kendall tests using the user-defined Matlab functions 'Mann_Kendall' and 'Mann_Kendall_Modified' available from https://uk.mathworks.com/matlabcentral/fileexchange/25531-mann-kendall-test and https://uk.mathworks.com/matlabcentral/fileexchange/25533-mann-kendall-modified-test, respectively.